\documentstyle[epsfig,amssymb,aps,prb]{revtex}

\begin{document}

\title{Magnetic field dependence of the energy of negatively charged
excitons in semiconductor quantum wells.}
\author{C. Riva\cite{clara} and F. M. Peeters\cite{francois}}
\address{Departement\ Natuurkunde, Universiteit Antwerpen (UIA),
Universiteitsplein 1, B-2610 Antwerpen, Belgium.}
\author{K. Varga\cite{kvarga}}
\address{Solid State Division, Oak Ridge National Laboratory, Oak Ridge, Tennessee 37831-3062}
\date{\today}
\maketitle

\begin{abstract}
A variational calculation of the spin-singlet and spin-triplet
state of a negatively charged exciton (trion) confined to a single
quantum well and in the presence of a perpendicular magnetic
field is presented. We calculated the probability density and the
pair correlation function of the singlet and triplet trion states.
The dependence of the energy levels and of the binding energy on
the well width and on the magnetic field strength was
investigated. We compared our results with the available
experimental data on GaAs/AlGaAs quantum wells and find that in
the low magnetic field region ($B<18$ T) the observed transition
are those of the singlet and the dark triplet trion (with angular
momentum $L_z=-1$), while for high magnetic fields ($B>25$ T) the
dark trion becomes optically inactive and possibly a transition
to a bright triplet trion (angular momentum $L_z=0$) state is
observed.

\end{abstract}

\nopagebreak
\section{Introduction}

After the initial work by Lampert,\cite{Lampert} who proved the
stability of the charged exciton complexes,  charged excitons in
bulk semiconductors \cite{Stebe75} as well as in an exactly
two-dimensional (2D) configuration \cite{Stebe89} were studied
theoretically. These studies revealed that, due to the
confinement, the 2D charged excitons have binding energies which
are an order of magnitude larger than charged excitons in the
corresponding bulk materials. The increased binding energy in
reduced dimensionality systems together with the improved
experimental techniques have allowed the experimentalists to
observe them in quantum well
structures.\cite{Kheng,Shields95,Finkelstein} Many of the
experimental results reported in the literature are for charged
excitons in the presence of a perpendicular magnetic
field.\cite{Finkelstein,Glasberg,manus1,Kim,Vanhoucke,Shields} Up
to recently, there was little or no agreement between  the
experimental results and the available
theories.\cite{Whittaker,Stebe2000}

Lately, however, progress was made in the direction of bringing
theoretical prediction and experiments closer to each other.
St\'eb\'e and Moradi\cite{Stebe2000} used a variational method
which was valid in the low magnetic field regime and explained
the minimum around $1$ Tesla observed experimentally by Shields
{\it et al.}\cite{Shields} in the charged exciton singlet
transition energy for a $300$ \AA \ wide quantum well. Recently
Muntenau {\it et al.} \cite{Mantenau} found a transition between
the singlet ground state and the triplet ground state at $B=35$ T
for a $200$ \AA \ wide asymmetric quantum well, similar to the
one predicted earlier by Whittaker and Shields\cite{Whittaker}
for a $100$ \AA \ wide symmetric quantum well.

The triplet transition energies which have been so far identified
are assigned to the angular momentum $L_z=-1$ triplet state. In
exactly 2D systems with translational invariance this state was
shown\cite{Dzyub1,Dzyub2} to be an optically {\it dark} state. As
a consequence, one would expect that such a state is 'dark' also
in quasi-2D systems, particularly in narrow quantum wells. The
fact that the $L_z=-1$ triplet is observed in quantum wells
suggests that a breaking of symmetry occurs and in particular
that the system is no longer invariant under a magnetic
translation. Recently, the existence of a bound {\it bright}
triplet state, i.e.  $L_z=0$ was predicted.\cite{wojsz} Due to its
small binding energy, this triplet state could be difficult to
detect. The possible existence of such a triplet state may force
us to review the assignment that has been made of some of the
photoluminescence lines.

Our previous works\cite{ascona,riva2} on charged excitons in
quantum wells was limited to the case of zero magnetic field and
showed that the stochastic variational method (SVM) is an
efficient technique for solving the effective mass Hamiltonian of
exciton complexes without involving any approximations. In Ref.
\onlinecite{riva2} we showed that approximations made by
St\'eb\'e {\it et al.}\cite{stebe} in the Coulomb matrix elements
lead to an overestimation of the trion binding energy. The latter
approximation aimed to convert the problem into an effective 2D
problem. In our approach no simplifying approximations are made
and the full 3D nature of the quantum well problem is retained.
Here we extend our previous work to the important experimental
situation in which an uniform magnetic field is applied along the
quantum well growth axis. Our results for the magnetic field
dependence of the trion singlet binding energy agrees, for the
first time, with available experimental results on 100 \AA \ and
300 \AA \
 wide GaAs/AlGaAs quantum wells. Furthermore we find
that the earlier predicted bright triplet is unbound for the 300
\AA \ wide quantum well and probably marginally bound for the 100
\AA \ wide quantum well.

The present paper is organized as follows. In Sec. II we present
the Hamiltonian of the problem and outline our method to obtain
the energy of the exciton and charged exciton. The conditional
probability density function of the trion, its pairs correlation
functions and the average distance between the different
particles in the trion are discussed in Sec. III. In Sec. IV we
compare our results for the transition energy and in Sec. V for
the binding energy with available experimental data on symmetric
GaAs/AlGaAs quantum wells and with the theoretical results of
Whittaker and Shields.\cite{Whittaker} In the last section we
summarize our results and present our conclusions.

\section{The model}

In the effective mass approximation  the Hamiltonian describing a
negative charged exciton, i.e. X$^{-}$, in an uniform magnetic
field B is given by
\begin{equation}
H=\sum_{i=1}^{3}{\frac{1}{2m_{i}}}({\vec{p}}_{i}-{\frac{e_{i}}{c}}{\vec{A}}
_{i})^{2}+\sum_{i=1}^{3}V({\vec{r}}_{i})+\sum_{i<j}{\frac{e_{i}e_{j}}{%
\varepsilon |{\vec{{r}}}_{i}-{\vec{{r}}}_{j}|},}
\end{equation}
where ${\vec{A}}_{i}={\frac{1}{2}}{\vec{r}}_{i}\times {\vec{B}}$
is the vector potential; $m_{i},e_{i}$ are the masses and charges
of the interacting particles; $\varepsilon$ is the dielectric
constant; the confinement potential is $V({\mathbf{r}}_i)=0$ if $
|z|<W/2$ and $V({\mathbf{r}}_i)=V_{i}$ if $|z|<W/2$, with $W$ the
quantum well width. The reference system is taken such that the
origin of the coordinate system is at the center of the quantum
well. For a GaAs/Al$_{x}$Ga$_{1-x}$As quantum well the heights of
the square well confinement potentials are $V_{e}=0.57\times
(1.155x+0.37x^{2})$ eV for the electrons and $V_{h}=0.43\times
(1.155x+0.37x^{2})$ eV for the hole. If we consider the case where
the magnetic field is applied along the growth axis of the well,
i.e. $ \vec{B}{=(0,0,B)}$, the Hamiltonian becomes:
\begin{equation}
H=\sum_{i=1}^{3}{\frac{1}{2m_{i}}}\left( -\hbar ^{2}\Delta
_{i}+{\frac{%
e_{i}^{2}B^{2}}{4c^{2}}}(x_{i}^{2}+y_{i}^{2})-{\frac{e_{i}\hbar B}{c}}%
l_{zi}\right)
+\sum_{i=1}^{3}V(\vec{r}_i)+\sum_{i<j}{\frac{e_{i}e_{j}}{%
|{\vec{{r}}}_{i}-{\vec{{r}}}_{j}|},} \label{ham}
\end{equation}
where $l_{zi}=-i\partial/\partial \phi_{zi}$ is the $z$-component
of the orbital momentum of the $i-$th particle. The Hamiltonian
under examination has cylindrical symmetry with respect to the
quantum well axis, i.e. $z$-axis, which implies that the
z-component of the total orbital angular momentum, $L_z$, is a
conserved quantity, i.e. a good quantum number. The spin
interaction is not explicitly included in our Hamiltonian. The
total spin of the electrons, $S_e$, and the spin of the hole,
$S_h$, and their projections along the z-axis, $S_{hz}$ and
$S_{ez}$, are conserved quantities. Notice that the state of the
system is not degenerate with respect to the total electron spin.
In fact the two electrons obey Fermi-Dirac statistics which means
that the electronic part of the total wave function must be
antisymmetric, i.e. when $S_e=0$ the spatial part of the
electronic wave function must be symmetric and when $S_e=1$ the
spatial part of the electronic wave function must be
anti-symmetric. Thus, $S_e$ can be used as a quantum number which
indicates the parity of the state. Once the projection along $z$
of the total orbital momentum, $L_z$, and the electron spin $S_e$
are fixed we obtain, after solving our Hamiltonian, a series of
energy levels which we indicate by the quantum numbers
$(n,L_z,S_e)$, where $n$ is the principal quantum number. These
levels are degenerate with respect to the quantum number
$S_h,S_{hz}$ and $S_{ez}$.

The Hamiltonian (\ref{ham}) is  solved using the stochastic
variational method which was outlined in Ref. \onlinecite{varga}.
The trial function, for the variational calculation, is taken as a
linear combination of ``deformed'' correlated Gaussian functions
(DCG),
\begin{equation}
\phi
_{N}(\vec{r}_{1e},\vec{r}_{2e},\vec{r%
}_{h})=\sum_{m=1}^{K}C_{qN}\Phi _{qN}(\vec{r}_{1e},
\vec{r}_{2e},\vec{r}_{h}), \label{wave-function}
\end{equation} with
\begin{equation}{ \small \Phi
_{qN}(\vec{r}_{1e},\vec{r}_{2e},\vec{%
r}_{h})={\cal A}\left\{\left(\sum_{r=1}^M\prod_{i=1}^{3}
\varphi_{qm_{ir}N}({ {\vec{\rho}}}_i) \right)  \exp \left[
-{1\over 2}\!\!\!\!\!\!\sum_{{\scriptsize
\begin{array}{c}
j,l\in \{1e,2e,h\} \\[-4pt]
 k\in \{x,y,z\}
 \end{array}}}\!\!\!\!\!\!\!\!\!\beta^{k}_{qjlN}(\vec{r}_{j}-\vec{r}_{l})^2\right]
\chi(1,2,3)\right\}, \nonumber }\end{equation}
 and
\begin{equation}
\varphi_{qm_{ir}N}(\vec{\rho}_i)=\xi_{qm_{ir}}(\vec{\rho_i})\exp\left(-\sum_{k\in
\{x,y,z\}}\beta^k_{qiiN}r^2_{ii}\right),
\end{equation}
where $r_{ik}$ gives the position of the $i$-th particle in the
$k$-direction; ${\cal A}$ is the antisymmetrization operator and
$\{C_{qN},\beta^k_{qilN}\}$ are the variational parameters,
$\chi(1,2,3)$ is the three particle spin function, and
$\xi_{qm_{ir}}({\bf\vec{\rho}})=(x+iy)^{m_{ir}}$ with $m_{ir}$
integers such that $L_z=m_{1r}+m_{2r}+m_{3r}$ for each value of
$r$, with $L_z$ the projection of the total angular momentum
along the $z$-axis; $M$ is the number of {\it channels} used to
obtain our state; $N$ indicates for brevity the set of quantum
numbers which characterizes our state, i.e. $(n,L_z,S_e)$. Note
that in contrast to the ``classical'' correlated Gaussians, here,
the parameter $\beta^k_{qjlN}$ which expresses the correlation
among the particle $j$ and the particle $l$ in the direction $k$,
is allowed to be different from the parameter $\beta^{k^{^\prime
}}_{qjlN}$ which couples the same two particles $j$ and $l$ in a
different direction $k^{^{\prime }}$. This additional degree of
freedom in the calculation allows us to take into account the
asymmetry introduced in the 3D space by the presence of the
quantum well and of the magnetic field.

 A basis of dimension $K$,
e.g. 10, is at first selected using the stochastic procedure. This
does not ensure that the best basis set is found, so a refinement
procedure is carried out on the basis set in order to improve it.
The refinement is made by replacing the $m$-th state with a new
state, i.e. with a state built using new parameters
$\{C_{mN},\beta^k_{milN}\}$ in such a way that the total energy is
lowered. When the refinement process does not change the total
energy significantly, the number of basis states is further
increased. The process is reiterated multiple times for different
and increasingly larger dimensions of the basis set, until the
energy reaches the desired accuracy. The final dimension of the
basis set consists typically of 400 states. Faster convergence is
obtained by taking into account the cylindrical symmetry, i.e. by
choosing $\beta^x_{qjlN}=\beta^y_{qjlN}$. Notice also that with
respect to the case without magnetic field, less basis states
have to be used because the magnetic field localizes the
particles around the magnetic center of mass leading to a faster
convergence of the energy. The number of channels used depends on
the magnetic field. For example, for the case $L_z=0$, we found
that for low magnetic fields we already obtain good results using
one channel, which actually gives the largest contribution, while
for large fields we have to use up to 7 channels, to obtain a
reasonable convergence. On the other hand, for small magnetic
fields we need larger number of states $K$ in order to accurately
describe the trion energy.

\section{Theoretical results}

Our numerical results are given for a GaAs/Al$_x$Ga$_{1-x}$As
quantum well. The parameter used in our calculation are $x=0.3$,
$\varepsilon =12.58$ and $m_{e}=0.067$ $m_{0},$ which give for
our unit of length  $a_{B }=\epsilon \hbar^2/e^2m_e=99.3$ \AA \
and energy $2R_{y }=e^2/\epsilon a_B=11.58$ meV. Notice that
$R_y$ and $a_B$ are calculated for the donor problem and do not
depend on the hole mass which we took to be  $m_h=0.34m_0$. Often
one uses $a^*_{B }=\epsilon \hbar^2/e^2\mu_e$ and $R^*_{y
}=e^2/\epsilon a^*_B$ where $\mu$ is the exciton reduced mass,
i.e. $1/\mu=1/m_e+1/m_h$, which for our problem is $\mu=0.056m_0$
corresponding to $a^*_B=118$ \AA \ and $R^*_y=4.8$ meV.

First we studied the magnetic field dependence of the
interparticle average distance. In Fig. \ref{average} we present
the 2D average distance, $d_{ij}=<\vec{\rho_{ij}}^2>^{1/2}$, vs.
the magnetic field for the electron-electron pair and for the
electron-hole pair, both in the $(n=0,L_z=0,S_e=0)$ state, i.e.
the singlet (solid curves) and in the $(n=0,L_z=-1,S_e=1)$ state,
i.e. the triplet (dashed curves) for a 100 \AA \ wide quantum
well. As a comparison we show also the exciton electron-hole
interparticle distance vs. magnetic field. For the exciton
problem the electron
 and the hole are more strongly bound and the interparticle
distance decreases more slowly than for the trion's singlet and
triplet state. Nevertheless, it decreases by $50\%$ over the
magnetic field range shown in the figure.
 For the negatively-charged exciton the
electron-electron average distance is always larger than the
electron-hole average distance both for the electron spin-singlet
state and for the electron spin-triplet state. This of course is a
consequence of the repulsive electron-electron interaction, while
the electron-hole is attractive. Notice that for $B=0$ the
electron-hole distance for the negative charged exciton is  about
twice the exciton one. The triplet state is more than 20 times
larger than the singlet-state in the small magnetic field range
where the triplet state is, in fact, unbound. The size of the
charged exciton decreases with increasing magnetic field. This
decrease is faster in the low magnetic field region, and it is
faster for the triplet than for the singlet state. The reason is
that the triplet state is more extended, it is less bound, and
consequently an external magnetic field will have a larger effect
on its size. Notice also that for both states, i.e. singlet and
triplet, the curves for $d_{ee}$ and $d_{eh}$ are almost parallel
to each other, but nevertheless with increasing magnetic field the
distance between them slowly decreases.

Next we calculated the 2D pair correlation function,
$g_{ij}^{2D}(\rho)=<\delta(|\vec{\rho}_i -\vec{\rho}_j|-\rho)>$,
for the spin-singlet and spin-triplet state of a charged exciton
in a quantum well of width 100 \AA \ in a magnetic field of
B$=13.7$ T, see Fig. \ref{corr}. We notice that the electron-hole
pair correlation function both for the spin-singlet state (dashed
curve) and for the spin-triplet state (dash-dotted curve) has its
maximum when the distance between the particles is zero. This
means that in both states the electron and hole have the tendency
of staying close to each other. Notice that the triplet
electron-hole pair has a longer tail compared to the singlet one,
indicating that the triplet is more extended but, nevertheless,
the particles in this state are still correlated even at large
distances. On the other hand the electron-electron pair
correlation function in the singlet state (solid curve) shows
that, even though the electrons have a significant probability of
being close to each other, the correlation is  maximal for
$\rho=0.35a_B$ which is a consequence of the Coulomb repulsion
between the electrons. In the triplet state the pair correlation
function is zero if  the particles are in the same position in
space, which is an expression of the Pauli exclusion principle,
and has a maximum at $\rho=1.32a_B$.

To gain further understanding on how the system is influenced by
the presence of a magnetic field, we studied the conditional
probability, which gives the probability of finding one of the
three particles in position $\vec{r}$ when the other two
particles are fixed at $\vec{r}_{1,0}$ and $\vec{r}_{2,0}$.
Notice that by fixing two of the particles we obtain information
on the positional correlation of the third particle. We focus on
the $\hat{xy}$-correlation since the effect of the applied
magnetic field along the quantum well axis is larger in the plane
orthogonal to the quantum well axis. Along the $z$-direction the
probability is mainly determined by the confinement potential.
Because the $x$- and $y$-axis are equivalent due to the
cylindrical symmetry of the problem we take $\vec{r}=(x,0,0)$ for
all three particles and for brevity we will indicate
$|\Phi(\vec{r},\vec{r}_{1,0},\vec{r}_{2,0})|^2$ by
$|\Phi(x,0,0)|^2$. In Fig. \ref{sing-ee}(a,b,c) we plot
$|\Phi(x,0,0)|^2$ for a negatively charged exciton in a 100\AA \
 wide quantum well when the two electrons are fixed at a distance
given by their average distance $d_{ee}=<\rho_{ee}^2>^{1/2}$.
Notice that for $B=0$ T (Fig. \ref{sing-ee}(a)) the hole is
centered around each of the two electrons, while for $B=13.7$ T
and for $B=54$ T (Fig. \ref{sing-ee}(b,c)) the hole is mostly
situated in the region between the two electrons. For $B=0$ T
there is a smaller but not zero probability that the hole is
between the two electrons. This binds the two electrons together.
When a magnetic field is applied the electrons are on the average
closer to each other and as a consequence the two "hole clouds"
around the electrons overlap. The hole has almost the same
probability of sitting on top of the two electrons or between
them. Notice, that when a magnetic field is applied the
conditional probability still shows two "kinks" at the position
of the two electrons, which are memories of the two peaks present
in the conditional probability function at $B=0$. Furthermore, for
increasing $B$ the hole wave function decays much faster when the
hole moves away from the electron. The increased probability for
the hole to sit between the two electrons leads to an increased
{\it bonding} between the electrons. This behavior is consistent
with the fact that the binding energy of the charged exciton
increases when a magnetic field is applied.

In Fig. \ref{sing-eh}(a,b,c) we plot $|\Phi(x,0,0)|^2$ for a
charged exciton in a 100\AA \ wide quantum well when the hole and
one electron are fixed at a distance equal to their average
position $d_{eh}=<\rho_{eh}^2>^{1/2}$, for the $B=0$ T case (Fig.
\ref{sing-eh}(a)), the $B=13.7$ T case (Fig. \ref{sing-eh}(b)) and
the $B=54$ T case (Fig. \ref{sing-eh}(c)). The qualitative
difference, between the situation when a large magnetic field is
applied and when a low magnetic field is applied is not very
pronounced, except for the length scale. However, we observe that
for $B=0$ T the probability of having the second electron near the
fixed electron is zero, while in the case in which a magnetic
field is applied there is a finite probability for the second
electron to be at the position of the first electron. Since the
charged exciton is in the singlet state, the spin function is
asymmetric for an interchange of the two electrons and
consequently there is no Pauli exclusion principle to forbid the
two electrons to be at the same position in space. Only the
electron-electron interaction will make the latter probability as
small as possible. This result is consistent with the result
obtained for the pair-correlation functions.

Next we consider the triplet state and limit ourselves to the
magnetic field $B=13.7$ T. Notice, that the triplet state is not
bound for small magnetic fields. We plot $|\Phi(x,0,0)|^2$ for a
charged exciton in a 100 \AA \  wide quantum well when the two
electrons are fixed (Fig. \ref{trip-ee}(a)) and when one electron
and the hole are fixed (Fig. \ref{trip-ee}(b)). Notice that there
is not much qualitative difference between the conditional
probability function of the triplet state and of the singlet state
(see Fig. \ref{sing-ee}(b)). Quantitatively there are two major
differences: i) the average distance between the particles is
substantially larger, and ii) the probability to find the second
electron at the same spatial position as the first one (see Fig.
\ref{trip-ee}) is zero, while this is not the case for the
singlet state. The latter is consistent with the fact that in the
triplet state the electronic part of the wave function is
antisymmetric under an exchange of the two electrons, which is
also consistent with the fact that the electron-electron
pair-correlation function is zero at the origin.

\section{Comparison of the transition energies with experiments}
In comparing our theoretical results with the available
experimental data we assume that the observed peaks in the PL
spectra are associated with an exciton, in which the electron and
the hole recombine with emission of light, or with  a
recombination of a negatively-charged exciton, which leaves
behind an electron in the lowest Landau level. Consequently, the
transition energies are defined as
\begin{eqnarray}
&E_{X}=E_g+E(X), \\ &E_{X^-}=E_g+E(X^-)-E_e(W,B),
\end{eqnarray}
where $E_g$ is the energy band gap and $E_e(W,B)$ is the energy of
a free electron in a quantum well of width W and in a magnetic
field of strength  $B$; $E(X)$ and $E(X^-)$ are, respectively, the
exciton and charged exciton total energy. We will also take into
account the Zeeman splitting induced by the magnetic field, under
the assumption that the transitions observed follow the energy
diagram discussed in Ref. \onlinecite{Vanhoucke}. We also assume
that the electron gyromagnetic factor, $g_e$, and the hole
gyromagnetic factor are the same for the exciton as well as for
the charged exciton. The total Zeeman splitting of each
transition can then be written, in agreement with the results
presented in Ref. \onlinecite{Shelling}, as
\begin{equation}
\Delta E_z=(g_e+g_h)\mu_BB, \label{Zeeman}
\end{equation}
where $\mu_B$ is the Bohr magneton. Notice also that the
gyromagnetic factor is defined using the same conventions as in
Ref. \onlinecite{Shelling}, i.e. the hole is considered to have
an effective spin of $\Sigma_h=1/2$ instead of the real hole spin
$S_h=3/2$. As a consequence of this Zeeman effect each transition
line $E$ is split into two lines, i.e. $E^{\pm}=E\pm\Delta
E_z/2$, associated to a change of 1 and -1 in the $z$-projection
of the total angular momentum $\vec{J}=\vec{L}+\vec{S}$, i.e.
$J_z$, respectively.

 In Fig. \ref{Shields-f} we compare our
theoretical  results for the transition energies of a $X^-$ in a
300 \AA \ wide quantum well (curves) with the experimental
results of Shields {\it et al.}\cite{Shields95,Shield-proc}
(symbols). We obtained the exciton gyromagnetic factor
$g_{ex}=g_e+g_h=1.16$ from the measured splitting between the
negatively ($\sigma_-$) and the positively ($\sigma_+$)
circularly polarized lines of Ref. \onlinecite{Shields95} using
Eq. (\ref{Zeeman}). This value of $g_{ex}$ is consistent with the
results by Ossau {\it et al.}\cite{Ossau} who found $g_{ex}=0.8$
for a 250 \AA \ wide quantum well. The experimental data presented
in Fig. \ref{Shields-f} are from the emitted negatively
($\sigma^-$) circular polarized light which results from
transitions with $\Delta J_z=-1$. We choose the energy gap such
that the exciton peak at $B=0$ T coincides with the experimental
exciton peak for B=0, which leads to $E_g=1521.55$ meV. Notice
that for the singlet we reproduce the experimental behaviour,
including the small minimum observed at low magnetic fields. Both
for the exciton and for the triplet state of the charged exciton
we find good agreement up to 8 T. At small magnetic fields: 1) the
theoretical results slightly overestimate the singlet transition
energy which is probably a consequence of the importance of
localization as argued, e.g., in Ref. \onlinecite{riva2}, and 2)
the triplet state is unbound for small magnetic fields and
consequently not observable. Notice also that the recently
discussed\cite{wojsz} bright triplet (dotted curve) is not bound
in the considered magnetic field region. None of the observed
transitions can be associated to such a bright triplet. The data
in the range 8-20 T are from Ref. \onlinecite{Shield-proc}
 and are obtained under different experimental conditions as compared to those
 from Ref. \onlinecite{Shields95} which were measured in the range 0-8 T.
 For example,
an increase in electron density will shift the experimental
photoluminescence towards larger energies.\cite{Huard} If we
perform an uniform shift of the experimental data by 0.5 meV in
the 8-20 T range, which leads to the open symbols, a much better
agreement with our theoretical results is obtained.

In Fig. \ref{Manus-f}(a,b) we compare our theoretical results for
the transition energies  of a 100 \AA \ wide GaAs/AlGaAs quantum
well with the experimental data obtained by Vanhoucke {\it et
al.}\cite{Vanhoucke} In Ref. \onlinecite{Vanhoucke} the Zeeman
splitting was measured to be $\Delta E_z/B=0.11$ meV/T leading to
$g_{ex}=1.85$ which is very different from the value $g_{ex}=0.1$
obtained in Ref. \onlinecite{Shelling} for a 115 \AA \ wide
quantum well. The energy gap is fixed by matching the $B=5$ T
experimental and theoretical $X^-$ singlet transition energies
which resulted into $E_g=$1520.35 meV. We use for the electron
and the hole mass $m_e=0.067 m_0$ and $m_h=0.34m_0$,
respectively. The lower transition line (squares) is in rather
good agreement with our results for the charged exciton singlet
transition energy. For $B<3$ T (see Fig. \ref{Manus-f}(a)) there
is a substantial deviation between theory and experiment which
again may be attributed to an enhancement of the negatively
charged exciton binding energy due to localization of the trion.
The higher transition line (circles) were attributed by the
authors of Ref. \onlinecite{Vanhoucke} to the triplet charged
exciton. Our theoretical results agree with this assignment at
least for $B< 20$ T (Fig. \ref{Manus-f}(a)). Notice that this
magnetic field range, i.e. $B<18$ T, is the same studied in Fig.
\ref{Shields-f} for the 300 \AA \ quantum well. In the high
magnetic field range (Fig. \ref{Manus-f}(b)), i.e. $B>25$ T, the
experimental results follow very closely the theoretical exciton
transition energy, which coincides practically with the $X^-$
bright triplet transition energy. In the intermediate magnetic
field range, i.e. $18$
 T$< B< 25$ T the results transit from the $X^-$ triplet to the
exciton transition or bright triplet transition.

From the above comparison we may construct the following picture:
1) in the magnetic field range $B< 18$ T quantum well width
fluctuations and disorder break the translational invariance of
the system which results into a breakdown of the optical
selection rule, thus allowing the dark triplet negatively charged
exciton state to be optically active; 2) only in the very small
magnetic field range, i.e. $B<5$ T,  the localization of the trion
due to quantum well width fluctuations leads to an increase of
the singlet and triplet $X^-$ binding energy. For the 300 \AA \
wide quantum well the effect of the quantum well width
fluctuations on the trion energy is substantially
smaller.\cite{riva2} This agrees with Fig. \ref{Shields-f} where
the magnetic field range over which the singlet binding energy is
strongly enhanced is much smaller, i.e. $B<2$ T, and the size of
the enhancement is also substantially smaller; 3) in the very
large magnetic field range $B>25$ T the optical selection rule is
restored and no transition from the $X^-$ dark triplet is
observed. Because of the inhibition of the decay  of the $X^-$
dark triplet it is possible that the bright triplet becomes
sufficiently populated making it experimentally observable. We
found that this $X^-$ bright triplet is at most marginally bound
 and therefore has almost the same transition energy as the exciton.

 For $B>40$ T the experimental results are slightly lower in
 energy as compared to our theoretical results. A possible reason
 for this deviation may be the importance of band non-parabolicity at such
 large magnetic fields. For example, if we increase the hole mass
 to $m_h=0.37m_0$ at $B=50$ T, the $X^-$ singlet  (exciton) transition energy becomes 1.5780 eV (1.5812 eV)
 which is almost 2 meV lower  than the $m_h=0.34m_0$ result 1.5796 eV (1.5824 eV), thus proving a strong
 dependence of the transition energy on the hole mass value. This is mainly due to the difference in confinement energy.  Notice, that the binding energy only changes from
 $2.8\pm0.1$ meV to $3.2\pm0.1$ meV, showing a less strong dependence on the hole mass.

 \section{ Comparison of the trion binding energy with experiments
and with other theoretical results}

 Finally we compute the binding energy of the
negatively charged exciton and compare it with the available
experimental results. The binding energy is defined as
\begin{equation}
E_B(X^-,B)=E(X)+E_e(W,B)-E(X^-),
\end{equation}
where $E(X)$ and $E(X^-)$ are respectively, the total energy of an
exciton and of a charged exciton in the quantum well and
$E_e(W,B)$ is the energy of a single electron in the quantum well
of width W.

In Fig. \ref{bindings} we present our results for  the binding
energy of a negatively charged exciton in a 300 \AA \ wide
GaAs/Al$_{0.3}$Ga$_{0.7}$As quantum well and we compare it with
the experimental binding energy obtained by Shields {\it et
al.}\cite{Shields95,Shield-proc} (symbols) and with the theory of
Whittaker and Shields\cite{Whittaker} (dotted and dash-dotted
curves). The error bars in the figure indicate the estimated
accuracy of our results. Note that the electron spin-singlet
binding energy (solid curve) increases with magnetic field, up to
about 35 T, after which it saturates. The electron spin-triplet
binding energy (dashed curves)  smoothly increases with magnetic
field up to 60 T. Notice the very good agreement between our
theory and the experimental binding energies both for the singlet
and triplet state up to about 13 T. For the lower magnetic field
range, $B< 2$ T, the binding energies are slightly underestimated
theoretically. We believe that the larger binding energy obtained
experimentally is a consequence of the localization of the trion,
as already noticed for the $B=0$ T case.\cite{riva2} The effect of
the magnetic field, however, decreases the discrepancy  between
theory and experiment. This is due to the fact that the magnetic
field increases the localization of the charged exciton, which is
then less sensitive to the well width fluctuations. In the range
$8$ T$ \leq B \leq 20$ T the experimental binding energies show
almost no magnetic field dependence which is in contrast to our
theoretical results which still increases with $B$, although less
fast than for $B \leq 8 $ T. As already mentioned the $8$ T$\leq B
\le 20$ T experimental results are measured under different
experimental conditions than those in the region $B\leq 8$ T.
Notice that our singlet binding energy is considerably larger than
the one obtained by Whittaker and Shields,\cite{Whittaker} while
the triplet binding energy is comparable to the one of Ref.
\onlinecite{Whittaker} up to 15 T. For $B> 15$ T the present
triplet binding energy becomes appreciably larger than the one of
Ref. \onlinecite{Whittaker}. One of the reasons for this
differences between our results and those of Whittaker and
Shields are the different parameters used in Ref.
\onlinecite{Whittaker}. They used $m_{h\parallel}^w=0.34m_0$,
$m_{e}^w=0.065m_0$ in the well, $m_{h\parallel}^b=0.45m_0$,
$m_{e}^b=0.07m_0$ in the barrier and $m_{h\perp}=0.18m_0$ in the
well and in the barrier, which partially explains the lower
binding energy.

 The binding
energy for a charged exciton in a 100 \AA \ wide
GaAs/Al$_{0.3}$Ga$_{0.7}$As quantum well is shown in Fig.
\ref{bindingm} and compared to the theory of Whittaker and
Shields\cite{Whittaker} (dotted and dash-dotted curves). Notice
that: i) we find substantial larger binding energies than
Whittaker and Shields,\cite{Whittaker} ii) no crossing between
the singlet and the triplet energies is found at least up to 70
T, while Whittaker and Shields predicted a singlet-triplet
crossing near 30 T, and iii) the bright triplet is at most
marginally bound for $B> 5$T. We find a binding energy of
$0.15\pm0.1$ meV while Wojs {\it et al.}\cite{wojsz} obtained a
binding energy of 0.75 meV for $B=20$ T (in Ref. \onlinecite{isa}
a reduced binding energy of 0.37 meV was reported). For the 300
\AA \ wide quantum well we found that the bright triplet state was
unbound for the considered magnetic field range.

The quantitative discrepancy between our theoretical results and
the one of Ref. \onlinecite{wojsz} is
 probably a
consequence of the approximations made by the authors of Ref.
\onlinecite{wojsz}: i) they replace the real quantum well $W$
with a hard wall quantum well with an effective width and only
the lowest subband is retained, ii) the 3D problem is replaced by
an effective 2D problem (in which the Coulomb interaction is
approximated by the 2D screened interaction: $e^2/\epsilon
\sqrt{\rho^2+\lambda^2}$), iii) the flat 2D quantum well geometry
is replaced by a Haldane sphere, and iv) only the lowest 5 single
particle Landau levels are included in their wave function.
Previously we showed\cite{riva2} for $B=0$, that the
approximations i) and ii) lead to an overestimation of the
binding energy of the charged exciton.\cite{riva2} Whittaker and
Shields \cite{Whittaker} showed that the inclusion of higher
subbands and of higher Landau Levels in the wave function
substantially increases the high field singlet binding energy,
while they have a smaller effect on the triplet binding energy.

 Note that in agreement with Whittaker and
Shields,\cite{Whittaker} and in contrast to the recent work by
St\'eb\'e and Moradi,\cite{Stebe2000} we find that the
spin-triplet state is unbound for $B=0$ T. This disagreement with
the work of St\'eb\'e and Moradi\cite{Stebe2000} can be traced
back to their poor variational function which gives an exciton
energy which is about $8\%$ larger than ours, while the
negatively charged exciton singlet energy is about $5\%$ lower
than ours.

 It has been argued that the hole mass is asymmetric and
that the in-plane hole mass depends on the magnetic field. One
expects that the hole mass in the z-direction, i.e. the
confinement direction, will almost not influence the exciton and
trion binding energies. This is different for the in-plane hole
mass which, e.g. through the reduced  exciton mass $\mu$, will
change the exciton and to a lesser extent the trion energies. In
a recent cyclotron  resonance experiment by Cole {\it et
al.}\cite{cole} on p-doped (311)A GaAs quantum wells the measured
hole mass varies from $m_h \approx 0.15-0.18 m_0$ for $B<5$ T to
$m_h\approx 0.35 m_0$ at higher fields for a 150 \AA\ wide
quantum well. For wider wells the large hole mass value was
reached at smaller magnetic fields and therefore, this mass
variation is expected not to be relevant for the 300 \AA \
sample. In order to investigate the influence of the value of the
in-plane hole mass on the trion singlet and triplet binding
energy we compare in Fig. \ref{bindingm} our results with those
for the asymmetric hole mass (thin solid and dashed curves in
Fig. \ref{bindingm}) in which the in-plane hole mass was reduced
to $m_h=0.18m_e$. Notice that: i) the singlet trion binding
energy is substantially reduced (about 0.5 meV); ii) the triplet
binding energy is practically not altered and coincides with the
Whittaker and Shields\cite{Whittaker} results for $B< 15$ T, and
iii) there is a singlet-triplet crossing at about 40 T. With this
smaller hole mass the exciton reduced mass is diminished by
$13\%$ leading to a lower exciton binding energy and also to an
increase of the trion total energies. This shifts the theoretical
curves in Fig. \ref{Manus-f} in such a way that an unrealistic
low band gap of 1518.3 meV has to be assumed in order to match
the experimental and theoretical $B\approx 5$ T trion transition
energies. Furthermore, the agreement between theory and
experiment is lost for $B>10$ T and the experimental trion
singlet energy for $B< 3$ T is now higher than the theoretical
curve which disagrees with the idea of an enhanced trion binding
energy in this low field region due to quantum well width
fluctuations. These findings argue against such a reduced hole
mass, even in the low magnetic field range.

It should also be noted that the use of a cyclotron mass in our
calculation may be questionable. In a cyclotron resonance
experiment, transitions between two Landau Levels are induced and
from the transition energy $\hbar \omega^*=E_1-E_0$ one defines
the cyclotron mass $m^*_c=eB/c\omega^*$, where $E_n$ is the energy
of the n-th Landau Level. Notice that such a definition only
corresponds  to the effective hole mass if the hole mass is
independent of the Landau Level. Furthermore, e.g., electric
subband crossings and polaron effects may invalidate such an
assignment. A further argument against the use of the low
magnetic field cyclotron hole mass published by Cole {\it et
al.},\cite{cole} is that those  results are for the (311) GaAs
plane while the experiments of Vanhoucke {\it et
al.}\cite{Vanhoucke} were performed on samples with quantum wells
in the (100) plane. It is well known that in the  latter
crystallographic direction, with increasing density or increasing
magnetic field, the hole mass very quickly reaches a value in the
$m_h\approx 0.3-0.5m_0$ range, the exact value depends on the
quantum well width (see for example Ref. \onlinecite{Goldoni}).
We believe that this argues in favor of the use of $m_h=0.34m_0$
in the important $B>4$ T magnetic field region as we did.

For the 100 \AA \ wide quantum well no experimental results on
the trion binding energy are available. Therefore, we show in
Fig. \ref{Manus-f1} the energy difference between the two
transition lines as measured in Ref. \onlinecite{Vanhoucke} and
compare them with: 1) the negatively charged exciton singlet
binding energy (solid curve), 2) the energy difference between
the negatively charged exciton  dark triplet and singlet (dashed
curve), 3) the energy difference between the negatively charged
exciton bright triplet and singlet (dotted curve). To be complete
we also show the negatively charged exciton bright triplet binding
energy. This figure nicely illustrates how in the low magnetic
field region, and more precisely in the range 6-18 T the
experimental results  are clearly not related to the binding
energy of the $X^-$ singlet state but rather to the difference
between the dark triplet state and the singlet state energy. In
the high magnetic field region, i.e. $B >25$ T, the experimental
results are closer to the singlet state binding energy and to the
energy difference between the bright triplet state and the
singlet state.

\section{Conclusion}

We presented a calculation of the lowest energy levels of the
negatively charged exciton spectrum in a quantum well and in the
presence of a magnetic field which is perpendicular to the
quantum well plane. Our approach is based on the stochastic
variational method in which the trion wavefunction is expanded in
deformed correlated Gaussian functions. The important correlation
between the particles is built in this wavefunction and therefore
such an approach is well suited for problems in which the binding
of the system is a pure consequence of the particle-particle
correlation as is the case for the trion. We do not observe any
spin-singlet/spin-triplet transition using the symmetric mass
approximation, however such  a transition is found for the 100 \AA
\ wide quantum well if we use the asymmetric hole mass
approximation (i.e. a substantially lower in-plane hole mass), in
agreement with what was predicted by Whittaker and
Shields.\cite{Whittaker} The singlet-triplet transition is found
to occur at about $40$ T, in contrast to the predicted $B=30$ T
reported in Ref. \onlinecite{Whittaker}. We have argued that at
such high magnetic fields the larger in-plane hole mass should be
used and consequently we believe that this transition should not
occur in reality for $B < 70$ T. Muntenau {\it et
al.}\cite{Mantenau} observed a spin-singlet/spin-triplet
transition in an asymmetric quantum well in which electrons and
holes are spatially separated. Such a singlet-triplet transition
is then of the same nature as the one predicted for spatially
separated charged donor systems.\cite{isa,Marmorkos,clara1}

 A comparison between our theoretical results and available experiments gives good agreement for the trion
singlet and triplet energy. Particular good agreement is achieved
with the experimental results of Shields {\it et
al.}\cite{Shields95,Shield-proc} on the 300 \AA \ quantum well.
For the results on the 100 \AA \ quantum well we find good
agreement for the trion singlet state while for the higher energy
transition we find for $B< 20$ T that the results agree with the
dark triplet transition, while for $B>25$ T this transitions
agrees more closely with the exciton transition energy or the
bright triplet energy. Because the latter two have, in this
magnetic field region, practically the same energy we are not
able to make any definite assignment for this transition line.

\section{Acknowledgment}

Part of this work is supported by the Flemish Science Foundation
(FWO-Vl), the `Interuniversity Poles of Attraction Program -
Belgian State, Prime Minister's Office - Federal Office for
Scientific, Technical and Cultural Affairs', the "Onderzoeksraad
van de Universiteit Antwerpen", and the Flemish-Hungarian
Cultural exchange program. K. Varga was supported by the U. S.
Department of Energy, Nuclear Physics Division, under contract
No. W-31-109-ENG-39 and OTKA grant No. T029003 (Hungary).
Discussions with M. Hayne, T. Vanhoucke, A. Dzyubenko and
correspondence with A. Wojs are gratefully acknowledged.

\begin{figure}
\caption{The 2D average interparticle distance vs. the magnetic
field for the exciton, and the singlet and triplet state of the
charged exciton in a quantum well of width 100
\AA.}\label{average}
\end{figure}

\begin{figure}
\caption{The 2D pair correlation function vs. the magnetic field
for the exciton and the spin-singlet and spin-triplet state of a
charged exciton in  a 100 \AA \ wide quantum well.}\label{corr}
\end{figure}

\begin{figure}
\caption{The projection on the x-axis of the conditional
probability for the charged exciton for $B=0$ T (a), $B=13.7$ T
(b) and  for $B=54$ T (c) in a quantum well of width 100 \AA. The
symbols represent the fixed electrons.}\label{sing-ee}
\end{figure}

\begin{figure}
\caption{The projection along the x-axis of the conditional
probability for the charged exciton, for $B=0$ T (a), $B=13.7$ T
(b) and for $B= 54$ T (c) in a quantum well of width 100 \AA. The
symbols represent the fixed electron and the hole.}\label{sing-eh}
\end{figure}

\begin{figure}
\caption{The projection on the x-axis of the conditional
probability function for the triplet state, when the electrons
are fixed (a) and when one electron and the hole are fixed (b), in
quantum well of width 100 \AA \ and for $B=13.7$ T. The symbols
represent the fixed particles.}\label{trip-ee}
\end{figure}

\begin{figure} \caption{Comparison between the experimental
and theoretical transition energies for charged excitons and
excitons in a 300 \AA \ wide quantum well. The open symbols are
the experimental results for $B> 8$ T shifted by 0.5 meV.}
\label{Shields-f}
\end{figure}

\begin{figure}
\caption{Comparison between the experimental and the theoretical
transition energies for charged excitons and excitons in a 100
\AA \ wide quantum well. For clarity, the low magnetic field
region (a) and the high magnetic field region (b) are shown
separately.} \label{Manus-f}
\end{figure}

\begin{figure}
\caption{The binding energy of a charged exciton in a 300 \AA \
wide quantum well compared to the experimental data of Shields
{\it et al.}\cite{Shields95,Shield-proc} and to the theoretical
results by Whittaker and Shields.\cite{Whittaker}}
\label{bindings}
\end{figure}

\begin{figure}
\caption{The binding energy of a charged exciton in a 100 \AA \
wide quantum well calculated using the symmetric hole mass
approximation (thick curves) and the asymmetric hole mass
approximation (thin curves). The results are compared to the
theoretical results by Whittaker and Shields.\cite{Whittaker}}
\label{bindingm}
\end{figure}

\begin{figure}
\caption{Comparison of the difference in energy between the upper
and lower $\sigma^-$ transition lines in Ref. 10 (symbols) with
our theoretical binding energy for the negative trion singlet
state (solid curve), the energy difference between our
theoretical dark triplet and singlet state (dashed curve) and the
energy difference between our bright triplet and singlet state
(dotted curve).} \label{Manus-f1}
\end{figure}

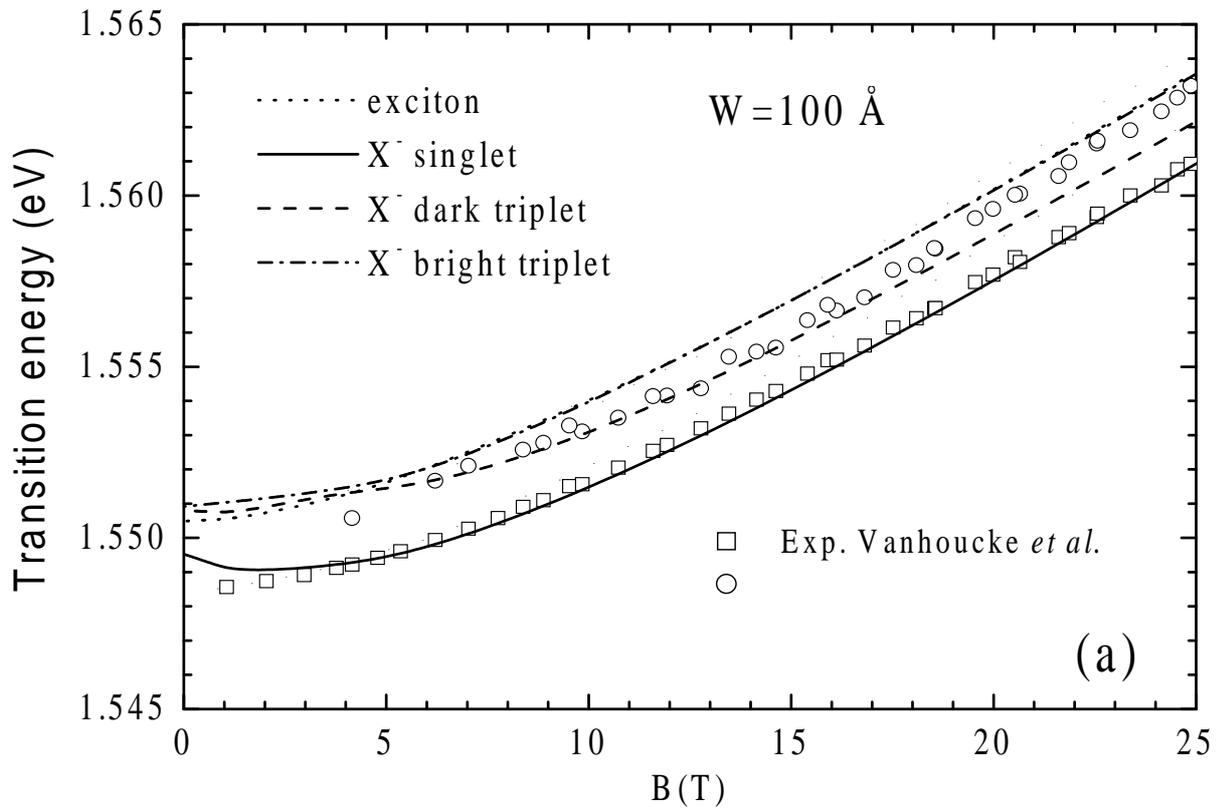
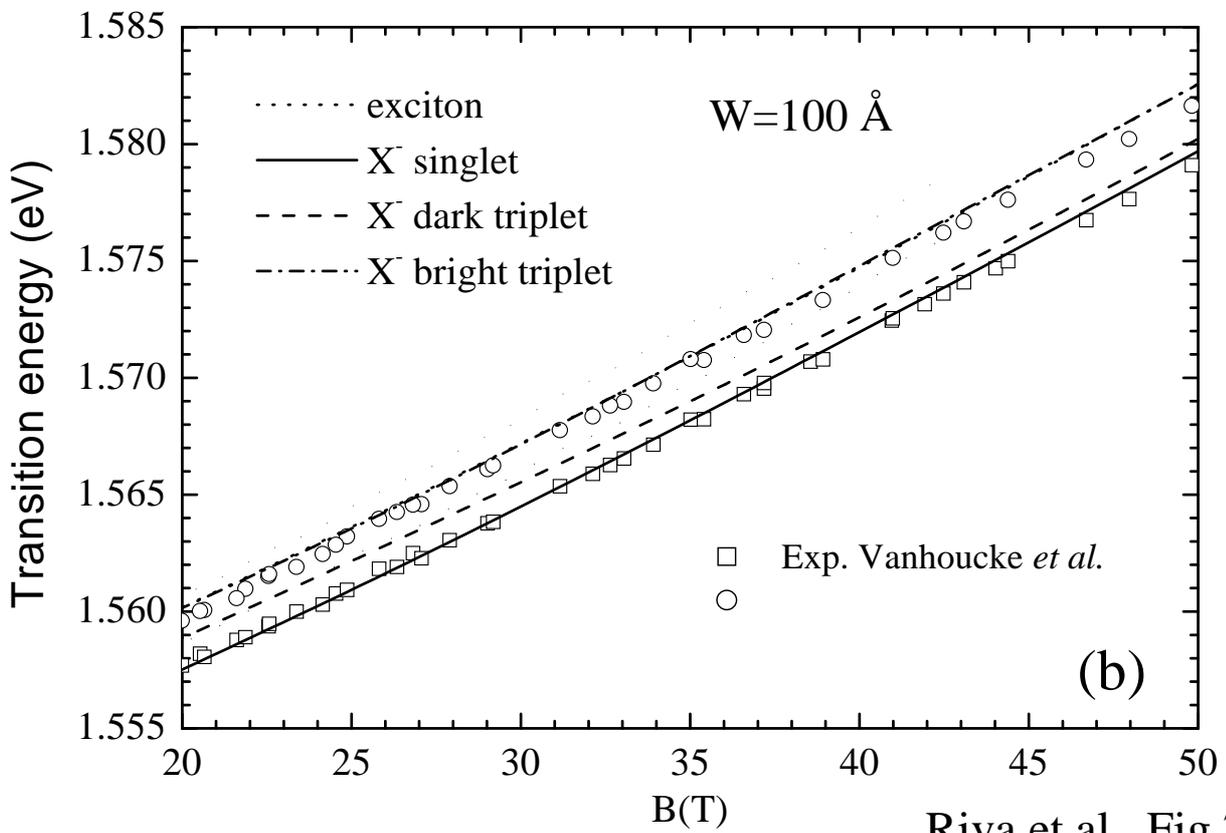
\end{document}